\documentclass[11pt,a4paper]{article}
\usepackage{jheppub_kim}
\usepackage{rotating} 
\usepackage{graphicx,epsfig}
\usepackage{amsmath}
\usepackage {amssymb}
\usepackage{subfigure}

\usepackage{relsize}

\providecommand{\U}[1]{\protect\rule{.1in}{.1in}}

\newcommand{\newc}{\newcommand}

\newc{\be}{\begin{equation}}
\newc{\ee}{\end{equation}}

\newc{\ba}{\begin{eqnarray}}
\newc{\ea}{\end{eqnarray}}
\newc{\bea}{\begin{eqnarray*}}
\newc{\eea}{\end{eqnarray*}}
\newc{\D}{\partial}
\newc{\ie}{{\it i.e.} }
\newc{\eg}{{\it e.g.} }
\newc{\etc}{{\it etc.} }
\newc{\etal}{{\it et al.}}
\newc{\lcdm}{$\Lambda$CDM }

\newc{\ra}{\Rightarrow}

\title{Modified cosmology though spacetime
thermodynamics and generalized mass-to-horizon entropy}
 \author[a,b,c]{Spyros Basilakos}
\author[d]{Andreas Lymperis}

\author[a,e]{Maria Petronikolou}

\author[a,f,g]{Emmanuel N. Saridakis}

\affiliation[a]{National Observatory of Athens, Lofos Nymfon, 11852 Athens, 
Greece}
\affiliation[b]{Academy of Athens, Research Center for Astronomy and Applied 
Mathematics, Soranou Efesiou 4, 11527, Athens, Greece}

\affiliation[c]{School of Sciences, European University Cyprus, Diogenes 
Street, Engomi, 1516 Nicosia, Cyprus}

\affiliation[d]{Department of Physics, University of Patras, 26500 Patras, 
Greece}

\affiliation[e]{Department of Physics, National Technical University of Athens, 
Zografou
Campus GR 157 73, Athens, Greece}

\affiliation[f]{Department of Astronomy, School of Physical Sciences, 
University of Science and Technology of China, Hefei 230026, P.R. China}

 \affiliation[g]{Departamento de Matem\'{a}ticas, Universidad Cat\'{o}lica del 
Norte, 
Avda.
Angamos 0610, Casilla 1280 Antofagasta, Chile}

\emailAdd{svasil@academyofathens.gr}
\emailAdd{alymperis@upatras.gr}
\emailAdd{msaridak@noa.gr}
\emailAdd{petronikoloumaria@mail.ntua.gr}

\abstract{ 
In this work we apply the  gravity-thermodynamics approach for the case of 
generalized mass-to-horizon entropy, which is a two-parameter extension of 
Bekenstein-Hawking entropy  that arises from the extended 
mass-to-horizon relation, that is in turn required in order to have consistency 
with the Clausius relation. We extract the modified Friedmann 
equations and we obtain  an effective dark energy sector arising from the 
novel terms. We derive analytical solutions for the dark energy density 
parameter, the dark energy equation-of-state parameter, and the deceleration 
parameter, and we show that the Universe exhibits the usual thermal history 
with 
the succession of matter and dark energy epochs. Additionally, depending on the 
value of the entropy parameters, the dark energy equation-of-state parameter 
can either  lie in the phantom regime at 
high redshifts entering into the quintessence regime at  small redshifts, 
or it can lie in the quintessence regime at high redshifts and   experience 
the  phantom-divide crossing at    small redshifts, while  in the 
far future  in all cases it 
asymptotically obtains the cosmological constant value $-1$. Finally, we 
perform   observational confrontation with Supernova Type Ia (SNIa), Cosmic 
Chronometers (CC)  and Baryonic Acoustic Oscillations   (BAO)   datasets, 
showing that the scenario  is in agreement with 
observations. 
}

\begin{document}
\maketitle

\section{Introduction}

Accumulated observational evidence  indicates that the Universe has undergone 
phases of accelerated expansion, both at early and late times. To account for 
this phenomenon, researchers typically follow two main approaches.  
The first approach involves developing new extensions and modifications of 
gravitational theories, essentially modifying the left-hand side of Einstein's 
field equations by incorporating additional terms into the standard 
Einstein-Hilbert action 
\cite{CANTATA:2021ktz,Capozziello:2011et,Cai:2015emx}. In particular, 
this avenue gives rise to modified gravity models, such as $f(R)$ gravity 
\cite{Starobinsky:1980te,Nojiri:2010wj}, 
$f(G)$ gravity \cite{Nojiri:2005jg, DeFelice:2008wz}, Lovelock gravity 
\cite{Lovelock:1971yv, Deruelle:1989fj}, Weyl gravity \cite{Mannheim:1988dj, 
Flanagan:2006ra},   Galileon gravity \cite{Nicolis:2008in, Deffayet:2009wt}, 
etc. Alternatively, one can work within the torsional formulation 
of gravity, and extending the teleparallel equivalent framework  to result  
to extensions such as $f(T)$ gravity \cite{Ben09, Linder:2010py, Chen:2010va}, 
$f(T,T_G)$ gravity \cite{Kofinas:2014owa,Kofinas:2014daa}, $f(T,B)$ gravity 
\cite{Bahamonde:2015zma}, teleparallel Horndeski theories 
\cite{Bahamonde:2019shr},
among others. Furthermore, one can start from the non-metric gravitational 
formulation 
and construct modifications such as  $f(Q)$ 
gravity \cite{BeltranJimenez:2017tkd,Heisenberg:2023lru}, $f(Q,C)$ gravity   
\cite{De:2023xua}, etc. 
The second approach retains general relativity as the underlying gravitational 
theory, and  modifies the right-hand side of Einstein's field equations 
by introducing novel matter fields, such as the inflaton field  or the 
 dark energy sector
\cite{Olive:1989nu,Bartolo:2004if,Copeland:2006wr,Cai:2009zp}. 
 
Apart from the aforementioned approaches, there 
exists a prominent conjecture suggesting that gravity  is deeply connected to 
thermodynamics \cite{Jacobson:1995ab,Padmanabhan:2003gd,Padmanabhan:2009vy}. 
Specifically, by treating the Universe as a thermodynamic system, composed of 
matter and dark energy and bounded by the apparent horizon 
\cite{Frolov:2002va,Cai:2005ra,Akbar:2006kj,Cai:2006rs}, one can
extract the Friedmann  by applying the first law of thermodynamics to the 
Universe horizon. This approach is known to work both in the framework of 
General Relativity, as well as in  the framework of  the various modified 
gravity theories of the literature, as long as one uses the corresponding 
entropy expression 
\cite{Cai:2006rs,Akbar:2006er,Paranjape:2006ca,Jamil:2009eb,
Cai:2009ph,Wang:2009zv,Jamil:2010di, 
 Gim:2014nba,Fan:2014ala}.

Nevertheless, there exists several generalizations of the standard entropy 
relation, which arise by abandoning the assumptions of the standard 
Boltzmann-Gibbs or Bekenstein-Hawking considerations. In these directions, 
abandoning the additivity assumption one obtains  R{\'e}nyi entropy 
\cite{renyi1961measures}, Tsallis entropy \cite{Tsallis:1987eu,Lyra:1998wz}, 
and 
Sharma-Mittal entropy \cite{sharma1975new}. Similarly, extending to 
relativistic 
statistical theory one obtains Kaniadakis   entropy 
\cite{Kaniadakis:2002zz,Kaniadakis:2005zk}, while incorporating quantum 
gravitational phenomena on the horizon one obtains Barrow entropy 
\cite{Barrow:2020tzx}. All these entropies  possess the standard 
 entropy as a particular limit of the involved parameter(s). Hence, the 
cosmological applications of these extended entropies in the framework of 
gravity-thermodynamics conjecture has attracted the interest of the community 
\cite{Lymperis:2018iuz,Saridakis:2020lrg,Nojiri:2019skr,
Nojiri:2019itp,Geng:2019shx,Lymperis:2021qty,Hernandez-Almada:2021rjs,
Zamora:2022cqz,Luciano:2022ely,Luciano:2022knb,Jizba:2022bfz,Dheepika:2022sio,
Luciano:2023zrx,
Luciano:2023fyr,Teimoori:2023hpv,Naeem:2023ipg,Jalalzadeh:2023mzw,
Basilakos:2023kvk, Naeem:2023tcu, Coker:2023yxr, 
Saavedra:2023rfq,Nakarachinda:2023jko,Jizba:2023fkp,
Okcu:2024tnw,Jalalzadeh:2024qej, Jizba:2024klq, Ebrahimi:2024zrk, 
Trivedi:2024inb,Okcu:2024llu, 
Petronikolou:2024zcj,Karabat:2024trf,Ens:2024zzs,Tsilioukas:2024seh,
Ualikhanova:2024xxe,Shahhoseini:2025sgl,Lymperis:2025vup,Nojiri:2025gkq} 
(see also \cite{Nojiri:2022dkr} where these entropies are generalized to a 
single multi-parametric form).

However, recently there is a discussion in the literature whether one can 
generalize the entropy relations without altering other thermodynamic 
relations \cite{Nojiri:2022sfd,Nojiri:2021czz,Gohar:2023lta}. In particular, in
\cite{Nojiri:2022sfd,Nojiri:2021czz} it is claimed that changing the entropy,  
exactly due to the first law of 
thermodynamics, requires a change in temperature or energy in order to have 
self-consistency. In similar lines, in 
\cite{Gohar:2023lta} it is argued that in order to have consistency with the 
Clausius relation and still use the Hawking temperature, then  a generalized 
mass-to-horizon relation must be introduced. This generalized 
mass-to-horizon relation leads in turn to a generalized entropy expression, 
which can encompass Tsallis-Cirto, Barrow, and other entropy types.

In this work we are interested in investigating the cosmological implications 
of the  generalized entropy that arises from the generalized mass-to-horizon 
relation. In particular, we apply the gravity-thermodynamics approach on the 
apparent horizon, incorporating the above generalized thermodynamic measures, 
and thus we result to modified Friedmann equations whose extra terms lead to an 
effective dark energy density and pressure. The plan of the work is as follows. 
In Section \ref{model} we review the standard  gravity-thermodynamics approach 
and then we apply it in the case of generalized mass-to-horizon entropy, 
extracting the modified Friedmann equations. In Section \ref{cosmicevolution} 
we analyze the corresponding cosmological evolution, focusing on the density 
parameters and the dark-energy equation-of-state parameters, as well as on 
confrontation with observations. Finally, Section \ref{concl} is devoted to the 
conclusions.

\section{Modified Friedmann equations through  generalized mass-to-horizon 
entropy} \label{model}

To begin our investigation, we first revisit the fundamental application of the 
first law of thermodynamics within the framework of General Relativity. We 
will then extend this approach by incorporating the generalized  
mass-to-horizon entropy in place of the conventional one. We consider an 
 Friedmann-Robertson-Walker (FRW)     metric
\begin{equation}
ds^2=-dt^2+a^2(t)\left(\frac{dr^2}{1-kr^2}+r^2d\Omega^2 \right),
\label{metric}
\end{equation}
  with $a(t)$   the scale factor, and where $k=0,+1,-1$ denotes   flat, 
close and open spatial geometry respectively. Additionally, 
 we consider that the Universe is filled with the matter 
perfect fluid, with energy density  and pressure $\rho_m$ and  $p_m$,  
respectively.

To implement the gravitational thermodynamics conjecture in a cosmological 
context \cite{Jacobson:1995ab,Padmanabhan:2003gd,Padmanabhan:2009vy}, the first 
law of thermodynamics is applied to the apparent horizon, which is given by
\cite{Bak:1999hd,Frolov:2002va,Cai:2005ra,Cai:2008gw}
\begin{equation}
\label{apphor}
r_a=\frac{1}{\sqrt{H^2+\frac{k}{a^2}}},
\end{equation}
  with $H=\frac{\dot a}{a}$ the Hubble function and where dots denote 
differentiation
with respect to cosmic time.
The next step is to attribute  an entropy and a temperature of the apparent 
horizon.  In the framework of General 
Relativity one applies the standard Bekenstein-Hawking entropy arising from 
black-hole thermodynamics \cite{Padmanabhan:2009vy},  
namely
\begin{equation}
\label{Horentropy}
S_{BH}=\frac{A}{4G} ,
\end{equation} 
with $A=4\pi r^{2}_{a}$     the horizon area and
$G$   the standard Newton's constant (throughout the work we use   natural 
units, i.e. we set $\hbar = k_{_B} = 
c =1$).
Furthermore,  
concerning the  horizon temperature we impose the standard relation   
  \cite{Gibbons:1977mu} 
\begin{equation}
\label{Th}
 T=\frac{1}{2\pi r_a}.
\end{equation} 

Now, the heat flow that crosses  the horizon during a time 
interval $dt$ can be calculated to be
$
\delta Q=-dE=A(\rho_m+p_m)H r_{a}dt  
$ \cite{Cai:2005ra}, while 
 the first law of thermodynamics  is as usual
$-dE=TdS$ (note that  we use the standard
 assumption that  the Universe fluid has the same temperature with the 
horizon due  to thermal equilibrium, which is expected to hold 
at late-time Universe
\cite{Padmanabhan:2009vy,Frolov:2002va,Cai:2005ra,Akbar:2006kj,
Izquierdo:2005ku,Jamil:2010di}.
 Differentiating the entropy relation  (\ref{Horentropy}) yields
$dS=2\pi r_a \dot{r}_a 
dt/G$, where $\dot{r}_a$ is easily calculated from  (\ref{apphor}). 
Thus, inserting  into the first law we obtain
\be
\label{cFE1}
-4\pi G (\rho_m +p_m)= \dot{H} - \frac{k}{a^2}.
\ee 
  Finally, inserting the matter conservation equation  
$
 \dot{\rho}_m +3H(\rho_m +p_m)=0 
$
 in (\ref{cFE1}) and integrating, we acquire
\be 
\label{cFE2}
\frac{8\pi G}{3}\rho_m =H^2+\frac{k}{a^2}-\frac{\Lambda}{3},
\ee
with $\Lambda$ the integration constant.

As we saw,  application of  gravity-thermodynamics conjecture on the Universe 
horizon does lead to the standard Friedmann equations. However, as we discussed 
above, if one repeats the same analysis but with generalized 
entropy relations, one will obtain modified Friedmann equations, and thus 
modified cosmological scenarios in which the extended entropies will give rise 
to an effective dark energy sector 
\cite{Lymperis:2018iuz,Saridakis:2020lrg,Nojiri:2019skr,
Nojiri:2019itp,Geng:2019shx,Lymperis:2021qty,Hernandez-Almada:2021rjs,
Zamora:2022cqz,Luciano:2022ely,Luciano:2022knb,Jizba:2022bfz,Dheepika:2022sio,
Luciano:2023zrx,
Luciano:2023fyr,Teimoori:2023hpv,Naeem:2023ipg,Jalalzadeh:2023mzw,
Basilakos:2023kvk, Naeem:2023tcu, Coker:2023yxr,
Saavedra:2023rfq,Nakarachinda:2023jko,Jizba:2023fkp,
Okcu:2024tnw,Jalalzadeh:2024qej, Jizba:2024klq, Ebrahimi:2024zrk, 
Trivedi:2024inb,Okcu:2024llu,
Petronikolou:2024zcj,Karabat:2024trf,Ens:2024zzs,Tsilioukas:2024seh,
Ualikhanova:2024xxe,Shahhoseini:2025sgl,Lymperis:2025vup,Nojiri:2025gkq}. 
Hence, in the following   we apply the  gravity-thermodynamics using 
the generalized mass-to-horizon entropy.

\subsection{ Generalized mass-to-horizon entropy}

As we mentioned in the Introduction, when one extends a thermodynamic quantity 
one should extend some of the other quantities too, in order to obtain 
self-consistency. In particular, since the  mass-to-horizon relation is one of 
the 
crucial ingredients in black-hole and Universe thermodynamics,  in 
\cite{Gohar:2023lta} the authors introduced a generalized mass-to-horizon 
relation, namely
\be 
M=\gamma \frac{c^2}{G}L^{n},
\label{Masstohor}
\ee
where $M$ is the mass and $L$ the cosmological horizon, and with 
$n$ a non-negative  constant and $\gamma$   a parameter with 
dimensions $[L]^{1-n}$. Inserting the above expression into the   Clausius 
relation $dE=c^{2}dM=TdS$,   requiring that $T$ is still given by the Hawking 
temperature (\ref{Th}), and considering that $L$ is the apparent horizon $r_a$, 
one results to a generalized entropy  given by \cite{Gohar:2023lta} 
 \be 
 \label{snentropy}
S_{n}=\gamma \frac{2n}{n+1}r_a^{n-1}S_{BH},
\ee
  where $S_{BH}$ is the standard Bekenstein-Hawking entropy (\ref{Horentropy}).
As one can see, in the case $\gamma=n=1$ one recovers both the standard  
mass-to-horizon relation, as well as the standard   Bekenstein-Hawking entropy.

\subsection{Modified cosmology through  generalized mass-to-horizon 
entropy}

In this section we desire to  extract the cosmological equations of the 
modified 
scenario, following the gravity-thermodynamics approach, but using the 
generalized  mass-to-horizon entropy (\ref{snentropy}) instead of the standard 
Bekenstein-Hawking one.  We start from  the first law of thermodynamics  
$-dE=TdS$, with $dE=A(\rho_{m}+p_{m})Hr_{A}dt$. We use   the standard Hawking 
 temperature (\ref{Th})   and the 
generalized entropy (\ref{snentropy}), which yields 
\be \label{snderivative}
dS_{n}=\gamma n\frac{2n}{G}r^{n}_{a}\dot{r}_{a}dt.
\ee
Substituting everything in the first law  we result to
\be \label{fried1}
-4\pi G(\rho_{m}+p_{m})=\gamma n\left (H^{2}+\frac{k}{a^2}\right 
)^{\frac{1-n}{2}}\left (\dot{H}-\frac{k}{a^2}\right ),
\ee
which is the second Friedmann equation. Therefore,  by integrating and  using 
the conservation equation $
\dot{\rho}_m +3H(\rho_m +p_m)=0$
we finally obtain
\be \label{fried2}
\frac{8\pi G}{3}\rho_{m}=\frac{2\gamma n}{3-n}\left (H^{2}+\frac{k}{a^2}\right 
)^{\frac{3-n}{2}}-\frac{\Lambda}{3}, 
\ee
where $\Lambda$ is the integration constant. Equation (\ref{fried2})  is the 
first Friedmann equation of the scenario at hand.

As we observe, application of the gravity-thermodynamics approach using the 
generalized mass-to-horizon entropy leads to modified Friedmann equations, and 
thus to modified cosmology. Focusing on a flat Universe  ($k=0$) we can 
re-write them in the standard form 
\begin{eqnarray}
\label{FR1}
&&H^2=\frac{8\pi G}{3}\left(\rho_m+\rho_{DE}\right)\\
&&\dot{H}=-4\pi G \left(\rho_m+p_m+\rho_{DE}+p_{DE}\right),
\label{FR2}
\end{eqnarray}
by introducing an effective dark energy density and pressure respectively as 
\begin{eqnarray}
&&
\!\!\!\!\!\!\!\!
\rho_{DE}=\frac{3}{8\pi G}\left [\frac{\Lambda}{3}+H^{2}-\frac{2\gamma 
n}{3-n}H^{3-n}\right ],
\label{rhode}\\
&& 
\!\!\!\!\!\!\!\!
p_{DE}=-\frac{1}{8\pi G}\left [\Lambda +\left (2\dot{H}+3H^{2}\right )-2\gamma 
nH^{1-n}\left (\dot{H}+\frac{3}{3-n}H^{2}\right )\right ].
\label{pde}
\end{eqnarray}
Finally, we can define the equation-of-state parameter for
the effective dark energy sector as
\be \label{wde}
w_{DE}\equiv \frac{p_{DE}}{\rho_{DE}} =-1-\frac{2\dot{H}\left (1-\gamma 
nH^{1-n}\right )}{\Lambda +3H^{2}-\frac{6\gamma n}{3-n}H^{3-n}}.
\ee

In summary,  applying the  gravity-thermodynamics approach with the generalized 
 
mass-to-horizon entropy leads to modified cosmology, with an effective 
dark-energy sector that depends on the entropy parameters. In the case  
$n=\gamma=1$, where standard entropy is recovered, the above scenario recovers 
standard $\Lambda$CDM paradigm.

\section{Cosmological behavior}
\label{cosmicevolution}

In the previous section we obtained a modified cosmological scenario  though 
spacetime thermodynamics and generalized mass-to-horizon entropy.
We can now proceed   to the extraction of analytical 
relations that provide the evolution of the Universe. 
Without loss of generality, we focus on the  dust matter case, setting 
 $p_m=0$. 
 
 \subsection{Cosmic evolution}
 
It proves convenient to introduce the matter and dark energy density parameters 
respectively as
 \begin{eqnarray} \label{omatter}
&&\Omega_m=\frac{8\pi G}{3H^2} \rho_m\\
&& \label{ode}
\Omega_{DE}=\frac{8\pi G}{3H^2} \rho_{DE}.
 \end{eqnarray} 
From the matter conservation equation , for dust matter we acquire $\rho_{m} = 
\frac{\rho_{m0}}{a^3}$, with $\rho_{m0}$ the value of the matter energy 
density at present scale factor $a_0=1$ (in the following the subscript ``0" 
marks the 
present value of a quantity). Thus, from   (\ref{omatter}) we immediately 
obtain 
$\Omega_m=\Omega_{m0} H_{0}^2/a^3 H^2$,  which though the relation  $\Omega_m + 
\Omega_{DE}=1$ gives
\be \label{h2}
H=\frac{\sqrt{\Omega_{m0}} H_{0}}{\sqrt{a^3 (1-\Omega_{DE})}}.
\ee 
In the following, for convenience, we will use the redshift $z$ as the 
independent variable, defined  as  $1+z=1/a$ for $a_0=1$. 
Differentiation of   (\ref{h2}) easily gives the useful expression
\be \label{hddot}
\dot H=-\frac{H^2}{2(1-\Omega_{DE})}[3(1-\Omega_{DE})+(1+z)\Omega'_{DE}],
\ee
where a prime denotes derivative with respect to $z$.

Substituting (\ref{rhode}) into (\ref{ode}) and using the relation(\ref{h2}) we 
 
obtain 
\be \label{omegade}
\Omega_{DE}(z)=1-\left \{\frac{3-n}{2\gamma n}\left  
(\sqrt{\Omega_{m0}}H_{0}\right )^{n-1}(1+z)^{\frac{3(n-1)}{2}}\left 
[1+\frac{\Lambda}{3\Omega_{m0}H^{2}_{0}(1+z)^3}\right ]\right 
\}^{\frac{2}{n-3}}.
\ee
This expression is the analytical solution for the dark energy density 
parameter  
$\Omega_{DE}(z)$, in a flat universe and for dust matter. Applying it at 
present 
 time, namely at $z=0$, we acquire
\be \label{lambdacons}
\Lambda =\frac{6\gamma n}{3-n}H^{3-n}_{0}-3\Omega_{m0}H^{2}_{0}.
\ee
 
Differentiating (\ref{omegade}) we obtain
\be \label{omegaprime}
\Omega^{'}_{DE}(z)=3\left(\mathcal{A} 
\mathcal{B}\right)^{\frac{2}{n-3}}(1+z)^{2}\left[-1-\frac{6H^{2}_{0}\Omega_{m0}
(1+z)^{3}}{\mathcal{B}(n-3)}\right],
\ee
with
\begin{eqnarray}\nonumber
\!\!\!\!\!\!\!\!\!\!\!
\mathcal{A}
&=&-\frac{(n-3)\left(H_{0}\sqrt{\Omega_{m0}}\right)^{n-3}}{6\gamma n}, 
\\ 
\nonumber
\!\!\!\!\!\!\!\!\!\!\!\!\!\!\!\!\!\!\!\!
\mathcal{B}&=&\Lambda +3H^{2}_{0}\Omega_{m0}(1+z)^3.
\end{eqnarray}
Furthermore, inserting (\ref{hddot}), (\ref{omegade}) and (\ref{omegaprime}) 
into equation  (\ref{wde}), we calculate  the other important observable, 
namely the dark-energy equation-of-state parameter as
\be \label{wdez}
w_{DE}(z)=-1-\frac{{18}\left(\mathcal{A} 
\mathcal{B}\right)^{-\frac{2}{n-3}}\left(1-\gamma 
n\mathcal{C}^{1-n}\right)H^{4}_{0}\Omega^{2}_{m0}(1+z)^{3}}{\mathcal{B}
(n-3)\left[\Lambda -\frac{6\gamma n\mathcal{C}^{3-n}}{3-n}+3\left(\mathcal{A} 
\mathcal{B}\right)^{-\frac{2}{n-3}}H^{2}_{0}\Omega_{m0}\right]},
\ee
where
\begin{eqnarray}\nonumber
\!\!\!\!\!\!\!\!\!\!\!\!\!\!\!\!\!\!\!\!
\mathcal{C}&=&\sqrt{\frac{H^{2}_{0}\Omega_{m0}}{\left(\mathcal{A} 
\mathcal{B}\right)^{\frac{2}{n-3}}}}.
\end{eqnarray}
Lastly, it proves convenient to introduce the decelaration parameter $q\equiv 
-1-\frac{\dot H}{H^2}$, which can be easily shown to be of the form
\be \label{qpar}
q(z)=\frac{1}{2}+\frac{3}{2}\Omega_{DE}w_{DE},
\ee
with $\Omega_{DE}$ and $w_{DE}$ 
  given by (\ref{omegade}) and (\ref{wdez}) respectively.
 In summary, we were able to extract analytical expressions for the important 
observable quantities. Hence, we can use them in order to  investigate in more 
detail the cosmological behavior in the scenario at hand.

\begin{figure}[!]
\centering
\includegraphics[width=6.cm]{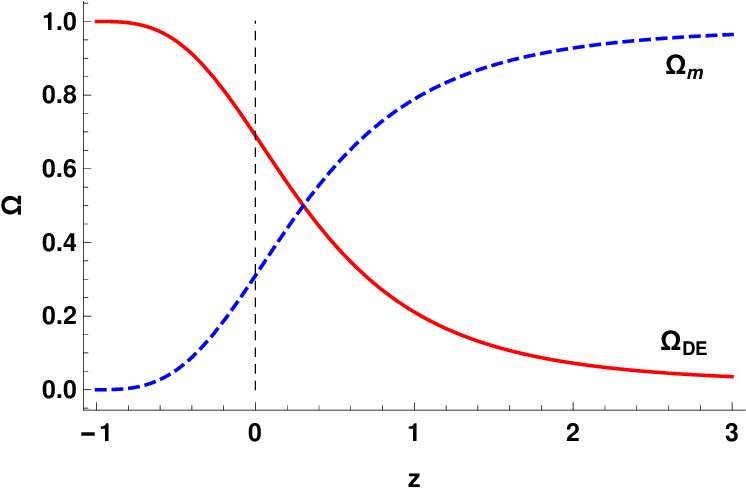} \\                                    
 \includegraphics[width=6cm]{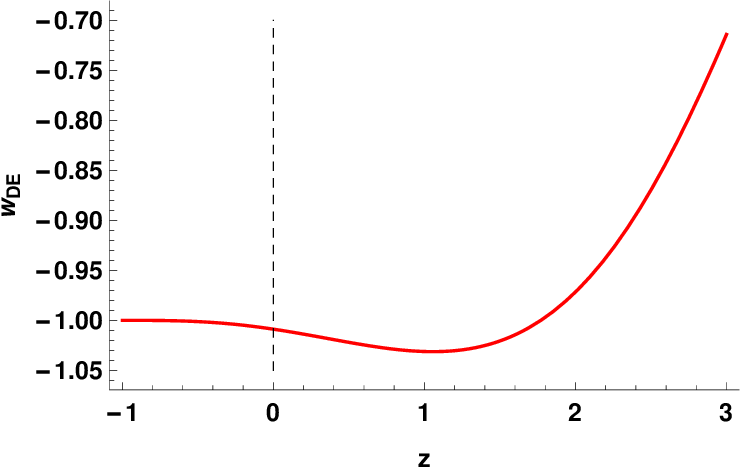} \\
\includegraphics[width=6cm]{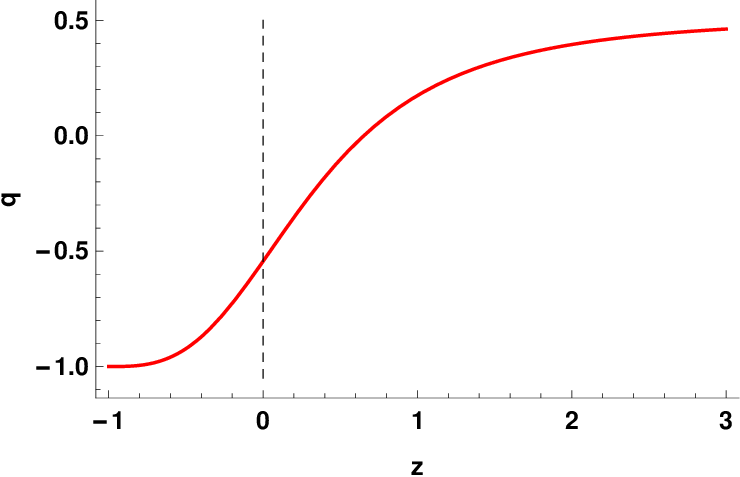}
\caption{\it{
\it{{\bf{Upper   graph}}: 
The evolution of the  effective dark 
energy density parameter $\Omega_{DE}$} (red-solid) according to 
(\ref{omegade}) and of the
matter density parameter $\Omega_{m}$ (blue-dashed) respectively, as a 
function of the redshift $z$, for the
modified cosmology through   generalized mass-to-horizon   entropy, in the 
case of  $\gamma =1$ (in units of $8\pi G=1)$ and $n=1.02$.
    {\bf{ Middle graph}}: 
The evolution of the corresponding  effective dark energy equation-of-state 
parameter $w_{DE}$ according to (\ref{wdez}). {\bf{ Lower   graph}}: 
The evolution of the   
deceleration parameter $q$   according to (\ref{qpar}). 
In all graphs we have imposed 
$\Omega_{DE}(z=0)\equiv\Omega_{DE0}\approx0.69$ at present,  and we 
have added a vertical dotted line denoting 
the current time $z=0$. 
}}
\label{fig:fig1}
\end{figure}

In the upper graph of Fig. \ref{fig:fig1} we present the evolution of the 
dark-energy  density parameter $\Omega_{DE}(z)$ according to (\ref{omegade}) 
and 
of the matter density parameter $\Omega_{m}(z) = 1-\Omega_{DE}(z)$, imposing 
the condition
$\Omega_{m0}\approx0.31$ \cite{Aghanim:2018eyx}.
 Additionally, in the middle graph we depict the 
corresponding evolution of the
dark-energy equation-of-state parameter $w_{DE}(z)$ according to (\ref{wdez}). 
Lastly, in the lower panel we present the evolution of the deceleration 
parameter  according to (\ref{qpar}). 
As we observe, in the  modified cosmological scenario at hand we can recover 
  the sequence of matter and dark energy eras, while in the future the Universe 
results to the      complete 
dark-energy dominated, de-Sitter phase. Note that  the 
transition 
from deceleration to acceleration is realized at $z\approx 0.6$ as 
required by observations. Concerning  the effective dark-energy 
equation-of-state parameter, we observe that   
its current value is   close $-1$ in agreement with observations, however in 
the 
past it exhibits an interesting dynamical behavior, experiencing the 
phantom-divide crossing, while in the future it asymptotically acquires the 
cosmological  constant value $-1$.

\begin{figure}[!]
\centering
\includegraphics[width=8cm]{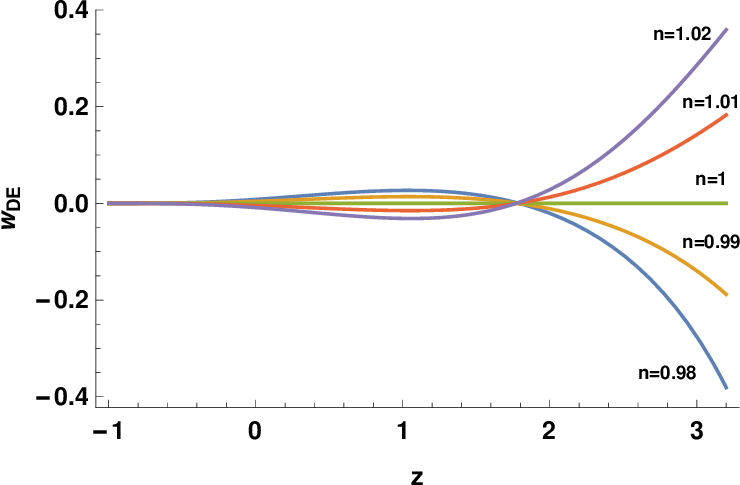} \\                                   
  
\caption{\it{
The  evolution of the effective dark energy equation-of-state 
parameter $w_{DE}$, for  $\gamma =1$ and  different values of the model  
parameter $n$. In all curves we have 
obtained density parameters 
behavior similar to the graphs of Fig.\ref{fig:fig1}, and we have imposed 
$\Omega_{DE}(z=0)\equiv\Omega_{DE0}\approx0.69$. 
}}
\label{fig:fig2}
\end{figure}

We proceed to examining the effect of the model parameters on the dark-energy 
equation-of-state parameter. In Fig. \ref{fig:fig2} we draw $w_{DE}$ for 
different values of  $n$. As we can see, for $n=1$ we recover $\Lambda$CDM 
cosmology as expected, however for $n$ deviating from 1 we obtain a very 
interesting behavior. In particular, we can see that for $n<1$ we acquire a 
phantom dark energy in the far past, while at small redshifts $w_{DE}$ enters 
into the quintessence regime, before resulting asymptotically to the 
cosmological-constant value from above. On the other hand,  for $n>1$ 
$w_{DE}$ lies in the quintessence regime in the far past, and exhibits the 
phantom-divide crossing at intermediate redshifts, before resulting 
asymptotically to the cosmological-constant value from below. We mention here 
that the dimensionless parameter $n$ does not need to deviate significantly 
from the standard value $n=1$ in order to obtain interesting cosmological 
behavior, which is an advantage since one expects that any realistic deviation 
from standard entropy should be small.

 \subsection{Observational constraints}

 In this subsection  we present the   observational constraints derived for 
the     modified cosmological scenario arising from the generalized 
mass-to-horizon entropy. We will use  data from Supernovae Type Ia (SNIa),
  Cosmic Chronometers (CC), and Baryonic Acoustic Oscillations   (BAO)  
measurements. In the following, $\phi^{\lambda}$ denotes the statistical 
vector  containing the free parameters of the scenario. 
 
Concerning   Supernovae Type Ia (SNIa), we utilize the binned Pantheon sample 
\cite{Scolnic:2021amr}, which approximates the full dataset with a reduced 
set $N = 40$ data points within the redshift range $0.01 \lesssim z 
\lesssim 1.6$.
The corresponding chi-square function is
  \begin{equation}
          \chi^{2}_{SN Ia}\left(\phi^{\nu}\right)={\bf \mu}_{\text{SNIa}}\,
          {\bf C}_{\text{SNIa},\text{cov}}^{-1}\,{\bf \mu}_{\text{SNIa}}^{T}\,,
          \end{equation}
where    ${\bf 
\mu}_{\text{\text{SNIa}}}=\{\mu_{1}-\mu_{\text{th}}(z_{1},\phi^{\nu})\,,\, 
...\,, \,   \mu_{N}-\mu_{\text{th}}(z_{N},\phi^{\nu})\}$. 
  The distance modulus can be expressed as $\mu_{i} = \mu_{B,i}-\mathcal{M}$, 
with 
$\mu_{B,i}$ the apparent magnitude at maximum brightness in the rest frame 
$z_{i}$. Additionally,   $\mathcal{M}$
  is introduced to account for the dependence of the observed 
  distance modulus, $\mu_{obs}$, on the fiducial 
cosmology \cite{Pan-STARRS1:2017jku}.
Finally, the  theoretical distance modulus reads as
\begin{equation}
\mu_{\text{th}} = 5\log\left[\frac{d_{L}(z)}{\text{Mpc}}\right] + 25,
\end{equation}
with $
d_L(z) = c(1+z)\int_{0}^{z}\frac{dx}{H(x,\phi^{\nu})}
$  
 the luminosity distance  assuming    flat geometry. 

 Concerning the Hubble constant observations,  
   we use the latest compilation of the $H(z)$ 
dataset    \cite{Yu:2017iju}, comprising of $N=31$     data points, 
within  $0.07 \lesssim z \lesssim 2.0$.
The corresponding $\chi^2_{H}$ function is expressed as 
  \begin{equation}
          \chi^{2}_{H}\left(\phi^{\nu}\right)={\bf \cal H}\,
          {\bf C}_{H,\text{cov}}^{-1}\,{\bf \cal H}^{T}\,,
          \end{equation}
with ${\bf \cal H 
}=\{H_{1}-H_{0}E(z_{1},\phi^{\nu})\,,\,...\,,\,
          H_{N}-H_{0}E(z_{N},\phi^{\nu})\}$ and where $H_{i}$ are the observed 
Hubble function values at  redshifts $z_{i}$ ($i=1,...,N$).

 \begin{table} 
    \centering
       \begin{tabular}{|c|c|c|}\hline
        \textbf{Parameters} & \textbf{CC+SnIa} & \textbf{CC+SnIa+BAO} \\ \hline
         $H_0$
         &  $69.0 \pm 1.8$ \,\text{km/s/Mpc}  & 
         $69.8 \pm 1.7 
$\,\text{km/s/Mpc}  \\
         $\Omega_{m0}$  &  $0.24 \pm 0.05$  &  $0.22 \pm 0.02$  \\
        $ n$  &  $1.08 \pm 0.06$  &  $1.09 \pm 0.01$  \\
        $ r_{\text{drag}}$  & $ 181.4 \pm 63.8 $ $\,\text{Mpc} $ & $  147.3 \pm 
1.6 $$\,\text{Mpc} $ \\ \hline
    \end{tabular}
    \caption{Observational constraints of cosmological and model parameters at 
$1\sigma$ confidence level.}
    \label{tab:Results}
\end{table}

Finally, concerning the  
Baryon Acoustic Oscillations (BAO) 
the sound horizon $r_d$, representing the maximum distance that  sound waves 
traveled before baryons decoupled in the early universe,  is given by 
\begin{equation}\label{soundhor}
    r_d = \int^{\infty}_{z_d}\frac{c_s(z)}{H(z)}dz ,
\end{equation}
where $z_d$ is the redshift of the drag epoch, while  the sound speed $c_s$ 
is 
\begin{equation}\label{soundspeed}
c_s(z) = 
\frac{c}{\sqrt{3\left(1+\frac{3\rho_{\beta}(z)}{4\rho_{\gamma}(z)}\right)}},
\end{equation}
where $\rho_{\beta}$ and $\rho_{\gamma}$ are the baryon and radiation energy 
densities respectively. 
The redshift of the drag epoch $z_d$ is 
\begin{equation}\label{zdrag}
    z_d=\frac{1925(\Omega_m h^2)^{0.251}}{1+0.659(\Omega_m 
h^2)^{0.828}}\left[1+b_1(\Omega_b h^2)^{b_2}\right],
 \end{equation}
    where $b_1$ and $b_2$ are factors given by
$
    b_1=0.313(\Omega_m h^2)^{-0.419}\times\left[1+0.607(\Omega_m 
h^2)^{0.6748}\right]$ and $    b_2=0.238(\Omega_m h^2)^{0.223}$.
We use the Sloan Digital Sky Survey (SDSS) and the DESI DR1 
BAO measurements obtained from various samples: The Bright Galaxy Sample (BGS, 
$0.1 < z < 0.4$), the Luminous Red Galaxy Sample (LRG, $0.4 < z < 0.6$ and $0.6 
< z < 0.8$), the Emission Line Galaxy Sample (ELG, $1.1 < z < 1.6$), the 
combined LRG and ELG Sample (LRG+ELG, $0.8 < z < 1.1$), the Quasar Sample (QSO, 
$0.8 < z < 2.1$) and the Lyman-$\alpha$ Forest Sample (Ly$\alpha$, $1.77 < z < 
4.16$). Finally, the chi-squared statistic used to fit the BAO data is
\begin{equation}\label{chisqBAO} \chi^2_{\text{BAO}}= (\Delta {X})^T 
\mathbf{C_{BAO}}^{-1} \Delta {X}, 
\end{equation}
where $\Delta {X} =  {x^{obs}} -  {x^{th}}$ and 
$\mathbf{C_{BAO}}^{-1}$ the inverse covariance matrix \cite{Chiang:2025qxg}.

\begin{figure*}[!]
    \centering
    \includegraphics[angle=-90,scale=0.55]{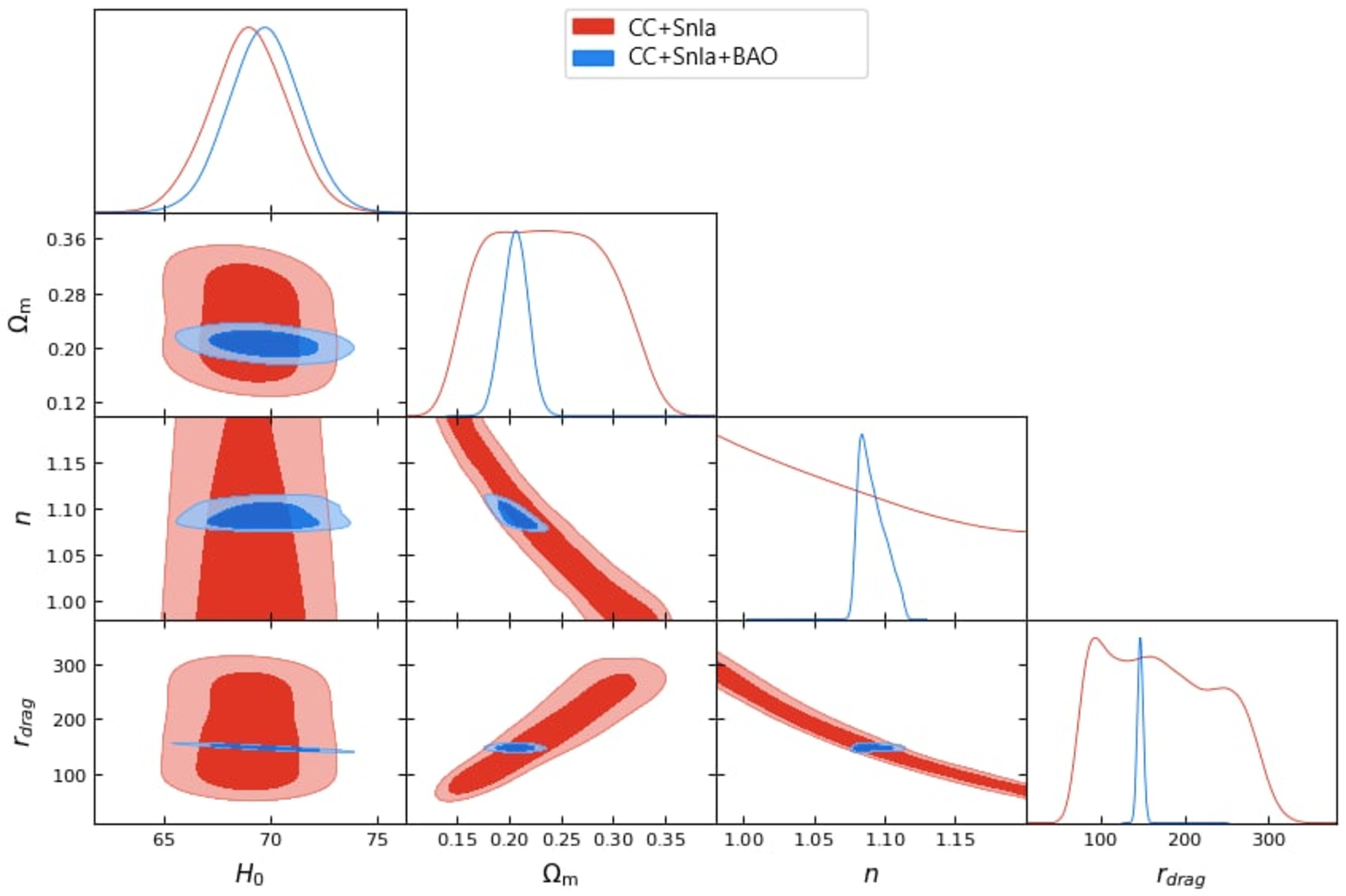}
    \caption{{\it{The $1\sigma$ and  $2\sigma$  iso-likelihood 
contours, for the 2D subsets of 
the parameter space  using  Cobaya code \cite{Torrado:2020dgo} (see text). 
 }} }
    \label{dataplot}
\end{figure*}

For our analysis we use the  Cobaya code 
\cite{Torrado:2020dgo}, which is publicly available.  In Table 
\ref{tab:Results} 
we depict the results on the parameters, while in Fig. \ref{dataplot} we show 
the corresponding  contour plots for 
the   model parameters. As we can see the scenario at hand is in agreement with 
observations. 

\section{Conclusions} \label{concl}

In this work we explored novel cosmological scenarios inspired by the 
well-established conjecture linking thermodynamics to gravity. Specifically, it 
is known that starting from the first law of thermodynamics applied at the 
Universe horizon, one can derive the Friedmann equations. The whole approach 
relies on the entropy relation, which is the standard Bekenstein-Hawking  one, 
as well as on the temperature relation, which is the Hawking one.
Hence, when various extended entropies are used, the same procedure leads to 
modified Friedmann equations and cosmological scenarios. 
However, when one modifies the entropy relation one should also change 
other thermodynamic quantities too, in order to obtain self-consistency. 
In particular,  it was shown recently that one such consistent approach is to 
change the mass-to-horizon ratio, and this, in turn, leads to a novel 
generalized 
entropy, characterized by two model parameters, namely $\gamma$ and $n$. 

We applied the gravity-thermodynamics approach with the generalized 
mass-to-horizon entropy expression, and we extracted modified Friedmann 
equations with an effective dark energy sector arising from the novel terms. In 
the case $\gamma=n=1$ where standard Bekenstein-Hawking entropy is recovered, 
the scenario at hand recovers $\Lambda$CDM paradigm, however in the general 
case  it exhibits integrating cosmological behavior.

Focusing on the flat case and considering dust matter enabled us to derive 
analytical solutions for the dark-energy density parameter, the dark-energy
equation-of-state parameter, and the deceleration parameter. Our 
analysis revealed that the Universe undergoes a transition from  
matter-dominated to dark-energy-dominated epochs, realized  at 
$z_{tr}\approx0.6$ in agreement with 
observations. Moreover, regarding the dark-energy equation-of-state parameter 
our analysis revealed that it is affected by the value of the  generalized 
mass-to-horizon parameter $n$, particularly in the past and present epochs. 
Specifically,   for $n<1$ we saw that dark energy lies in the phantom regime at 
high redshifts, entering into the quintessence regime at  small redshifts, 
 while in the future it  results asymptotically to   
$-1$   from above. Additionally,  for $n>1$ 
$w_{DE}$ lies in the quintessence regime at high redshifts and it experiences 
the  phantom-divide crossing at at  small redshifts, before 
resulting asymptotically to $-1$ from below in the far future. 
Finally, we performed a   confrontation with observations, using 
 Supernovae Type Ia (SNIa),
  Cosmic Chronometers (CC), and Baryonic Acoustic Oscillations   (BAO)  
datasets, showing that the scenario at hand is in agreement with 
observations.

In summary, the modified cosmological framework derived from the application of 
the
gravity-thermodynamics conjecture using the generalized 
mass-to-horizon entropy leads to interesting cosmological phenomenology. 
However, there are more investigations that need to be performed before one 
considers it as a successful candidate for the description of Nature. For 
instance, one should perform a detailed analysis at the perturbative level,
aiming to see how matter overdensity and   growth of structures evolve. 
Additionally, one should perform a complete dynamical-system analysis
in order to reveal the global behavior  of the Universe, independently off the 
specific evolution. These necessary investigations lie beyond the scope of the 
present work and are left for future projects.


\end{document}